\documentclass[onecolumn]{emulateapj}
\usepackage{graphicx}
\usepackage{amsfonts,amsmath,amssymb,mathrsfs}

\newcommand{\be}{\begin{eqnarray}}
\newcommand{\ee}{\end{eqnarray}}
\newcommand{\rar}{\rightarrow}

\begin{document}

\title{Constraining the quadrupole moment of stellar-mass black-hole candidates with the continuum fitting method}

\author{Cosimo Bambi}
\affil{Institute for the Physics and Mathematics of the Universe, 
The University of Tokyo \\ Kashiwa, Chiba 277-8583, Japan}
\email{cosimo.bambi@ipmu.jp}

\author{Enrico Barausse}
\affil{Department of Physics, University of Maryland \\ 
College Park, Maryland 20742, United States}
\email{barausse@umd.edu}

\date{\today}

\begin{abstract}
Black holes in General Relativity are known as Kerr black holes and 
are characterized solely by two parameters, the mass $M$ and the spin 
$J$. All the higher multipole moments of the gravitational field are
functions of these two parameters. For instance, the quadrupole moment 
is $Q=-J^2/M$, which implies that a measurement of $M$, $J$, and $Q$ 
for black hole candidates would allow one to test whether these objects 
are really black holes as described by General Relativity. While future 
gravitational-wave experiments will be able to test the Kerr nature 
of these objects with very high accuracy, in this paper we show that 
it is possible to put constraints on the quadrupole
moment of stellar-mass black hole candidates by using presently 
available X-ray data of the thermal spectrum of their accretion disk.
\end{abstract}



\section{Introduction}

The most general stationary and axisymmetric black-hole (BH) solution of
Einstein's equations in a four-dimensional, asymptotically flat spacetime 
is given by the Kerr geometry~\citep{Kerr:1963ud}.
Today there are at least two classes of astrophysical BH 
candidates: stellar-mass objects in X-ray binary systems 
(mass $M \sim 5 - 20$~$M_\odot$)~\citep{stmbhs1,stmbhs2} and 
super-massive objects at the center of most galaxies 
($M \sim 10^5 - 10^{10}$~$M_\odot$)~\citep{sumbhs}. 
The existence of a third class of objects, intermediate-mass BHs with  
$M \sim 10^2 - 10^4$~$M_\odot$~\citep{imbhs}, 
 is still controversial because their detections are indirect and 
definitive dynamical measurements of their masses are still lacking~\citep{imbhs}.

All these objects are supposed to be Kerr BHs because they 
cannot be explained otherwise without introducing new physics. 
In particular, stellar-mass BH candidates in X-ray binary 
systems are too heavy to be neutron or quark stars for any reasonable 
matter equation of state~\citep{ruf,kal}. Observations of stellar orbits around the super-massive BH 
candidate Sgr A$^\star$ at the center of the Galaxy show that this object is too massive, compact, 
and old to be a cluster of non-luminous bodies~\citep{maoz} or a fermion ball~\citep{2002Natur.419..694S} 
(\textit{i.e.,} an object made of sterile neutrinos, gravitinos
or axinos supported
by degeneracy pressure~\citep{1998ApJ...500..591T}). Other exotic alternatives
such as boson stars~\citep{boson} and gravastars~\citep{gravastar,gravastar2,gravastar3} 
seem to be disfavored by the near-infrared observations of Sgr A$^\star$~\citep{2006ApJ...638L..21B,2009ApJ...701.1357B}.

In spite of this body of indirect evidence, a definitive proof that BH candidates are really
described by the Kerr solution of General Relativity is still elusive. A framework within which to test the 
Kerr BH hypothesis was first put forward by~\citet{ryan_multipoles,ryan_3,ryan_4}, who considered a
general stationary, axisymmetric, asymptotically flat, vacuum
spacetime. Such a generic spacetime can be used to describe the gravitational field around a central object, whatever its nature, 
and its metric can be expressed in terms of the mass moments $M_\ell$
and current moments $S_\ell$~\citep{multipoles1,multipoles2}. Assuming reflection symmetry,
the odd $M$-moments and even $S$-moments are identically
zero, so that the non-vanishing moments are the mass $M_0=M$, the mass
quadrupole $M_2=Q$ and the higher-order even terms $M_4, M_6,
\ldots$, as well as the angular momentum $S_1=J$, the current octupole
$S_3$ and the higher-order odd terms $S_5, S_7, \ldots$. 
In the case of a Kerr BH, all the moments  $M_\ell$
and $S_\ell$ are locked to the mass and angular momentum
by the following relation:
\begin{equation}\label{kerrMultipoles}
M_\ell+{\rm i}S_\ell=M\left({\rm i}\frac JM\right)^\ell\;.
\end{equation}
This is the celebrated ``no-hair'' theorem \citep{hair1,hair2,hair3}: an (uncharged) stationary
BH is uniquely characterized by its mass and spin angular momentum. 
Therefore, a measurement of the mass, spin and higher moments (starting
with the quadrupole moment $Q$) of BH candidates would permit testing Eq.~(\ref{kerrMultipoles})
and therefore the Kerr-nature of these objects.

Ryan's idea was to use future gravitational-wave observations of extreme-mass ratio inspirals 
(EMRIs, \textit{i.e.,} systems consisting of a stellar-mass 
BH orbiting a super-massive BH in a galactic center) to perform this test. 
EMRIs will be a key source for the future-space based detector LISA: because the stellar-mass BH will
orbit the super-massive BH $\sim 10^6$ times during LISA's lifetime, slowing spiralling in due to the emission
of energy and angular momentum via gravitational waves, even a small deviation from the Kerr geometry will build
up an observable dephasing in the gravitational waveforms, thus allowing one to map the spacetime
of super-massive BHs with very high accuracy. Ryan's spacetime mapping idea 
originated a whole line of research aiming at using LISA's observations of EMRIs to test 
the Kerr nature of super-massive 
BHs~\citep{bumpy1,bumpy2,apostolatos1,apostolatos2,kostas,gair,kesden_boson,torus,hydrodrag,barack_cutler} and 
even General Relativity itself~\citep{nicoTestCS,comment_on_psaltis}. Another independent (and complementary)
test of the no-hair theorem with LISA uses BH quasi-normal modes~\citep{emanuele1}. Because the frequencies of these
modes encode the multipolar structure~(\ref{kerrMultipoles}) of the Kerr geometry, they can be used to test
consistency with the Kerr solution and to distinguish it from boson stars~\citep{emanuele2} or gravastars~\citep{gravastar2}.

Besides these tests based on gravitational waves, there are other
proposals using electromagnetic radiation. Constraints on the 
quadrupole moment of the compact companion of radio pulsars can be 
obtained with timing measurements~\citep{wex}. Astrometric monitoring 
of stars orbiting at milliparsec distances from Sgr A$^\star$ may be 
used to test the no-hair theorem for the super-massive BH candidate 
at the center of the Galaxy~\citep{will1,will2}. A very promising way 
to measure deviations from the Kerr metric is represented 
by future observations of the ``shadow'' of super-massive BH candidates 
through very long baseline interferometry (VLBI) experiments 
\citep{naoki,psaltis1,psaltis2}. The study of quasi-periodic variability 
in BH spectra may also test the geometry of the spacetime around
BH candidates~\citep{psaltis3}. 
Remarkably, \citet{psaltis4} also shows that the data for iron K$\alpha$ emission lines in thin accretion
disks can \textit{already} constrain deviations from the Kerr geometry.
Although these measurements yield much less accurate constraints than what will be achieved with LISA,
and can be subject to critiques (see~\citet{tita}, who show that iron-K$\alpha$ lines 
with the same features as those attributed to BH
candidates are observed also around white dwarfs), these data are available \textit{now}, which is not the case for all the other tests reviewed above.
However, because of the controversial interpretation of the origin of these lines, and because \citet{psaltis4} finds a degeneracy between 
the spacetime's quadrupole and spin (\textit{i.e.,}
similar shapes for the iron-K$\alpha$ lines can be obtained with a Kerr BH or with a non-Kerr object 
with spin and quadrupole slightly shifted from
the Kerr values), it is important to explore other techniques to test the no-hair theorem with \textit{present} data.

In this paper we propose using the continuum spectrum of BH candidates, which
has been shown to be potentially a promising tool to tell Kerr BHs from specific alternatives such as
gravastars~\citep{harko_gs}, BHs in Chern-Simons gravity~\citep{harko_cs}, 
BHs in Horava-Lifschitz gravity~\citep{harko_hl} or certain classes of
naked singularities~\citep{harko_ws,rohta,harko_ns}. While these attempts highlighted 
some important differences between the spectra of these objects and those of Kerr BHs,
they relied on specific models for the BH candidate, and did not investigate whether presently available data
allow one to break the degeneracy mentioned above between the parameters of these objects and those of a Kerr BH
(\textit{i.e.,} whether present X-ray data can tell the spectrum of a non-Kerr object from that of a Kerr BH with
arbitrary $J$ and $M$). In this paper we address both issues, \textit{(i)} by considering a very general model for the
BH candidate (\textit{i.e.,} one which allows its quadrupole moment
to slighlty deviate from the Kerr value, thus approximately describing a variety
of almost-Kerr objects), \textit{(ii)} by comparing our model to present X-ray data, although in a simplified way,
and \textit{(iii)} by discussing the sources of systematic error that might affect the data and that must be properly
understood before one can perform robust tests of the no-hair theorem.

In the range 0.1~keV -- 1~MeV, the 
generic spectrum of a stellar-mass BH candidate is 
characterized by three components, even if their relative intensities 
vary with the object and, for a given object, with time: $i)$ a soft 
X-ray component (energies $< 10$~keV), $ii)$ a hard power law X-ray 
component with an exponential cutoff (energies in the range 
$10-200$~keV, photon spectral index in the range $1-2.5$), and 
$iii)$ a $\gamma$-ray component (energies $> 300$~keV). For a review, 
see e.g. \citet{liang}. The soft X-ray component is commonly 
interpreted as the thermal spectrum of a thin disk, while the exact 
origin of the other two components is not so clear.

Geometrically thin and optically thick accretion disks can be 
described by the Novikov-Thorne model~\citep{n-t}. They are expected 
when the accretion flow is radiatively efficient, which requires
a luminosity $L \lesssim 0.3$~$L_{Edd}$, where $L_{Edd}$ is the
Eddington limit. The emission is blackbody-like. Assuming that the 
inner edge of the disk is at the innermost stable circular orbit 
(ISCO)\footnote{Such an assumption is supported either by 
observational facts~\citep{steiner} and numerical 
simulations~\citep{shafee,penna} (but see~\citet{krolik}).}, the disk luminosity of a Kerr 
BH is determined only by its mass, $M$, the mass accretion 
rate, $\dot{M}$, and the spin parameter, $a=J/M^2$. This fact can thus 
be exploited to estimate the spin of stellar-mass BH 
candidates~\citep{zhang}. This is the continuum fitting method and 
at present has been used to estimate the spin parameter of a few 
stellar-mass BH candidates~\citep{mcclintock}\footnote{For 
super-massive BHs, the disk temperature is lower (the 
effective temperature scales like $M^{-0.25}$) and this approach 
cannot be applied.}. Basically, knowing the mass of the object, 
its distance from us, and the inclination angle of the disk, it 
is possible to fit the soft X-ray component of the source and 
deduce $a$ and $\dot{M}$.

In this paper, we compute the thermal spectrum of a geometrically
thin and optically thick accretion disk around a generic compact
object. We use a subclass of Manko-Novikov spacetimes~\citep{m-n}, 
which are stationary, axisymmetric, and asymptotically flat 
exact solutions of the vacuum Einstein equations. 
In addition to the mass and the spin of the massive 
object, here we have the anomalous quadrupole moment, $q$. The 
latter measures the deformation of the massive object with respect 
to a Kerr BH: when $q>0$, the object is more oblate than 
a Kerr BH, when $q<0$, it is more prolate, while, for $q=0$, 
we recover the Kerr metric. The value of $q$ determines the radius
of the ISCO and changes the high frequency region of the spectrum 
of the disk.

In general, this makes the spectrum of the disk almost degenerate
in $a$ and $q$. However, only in the 
Kerr case the radius of the ISCO goes to $M$ as $a$ approaches 1. 
For $q\neq0$, even a small deviation from the Kerr metric makes
the radius of the ISCO significantly larger than $M$. Since current
X-ray observations suggest that there are objects with small ISCO
radius, one can in principle obtain 
interesting constraints on the value of $q$. 

The purpose of this paper is therefore to present a preliminary
investigation, showing that X-ray continuum spectra can potentially 
be used to constraint small quadrupole deviations away from
the Kerr metric, once all the physical effects have been
included in the model and all systematics have been understood.
In particular, our computation of the disk's spectrum does not
include the effect of light bending. This is a simplification
of our model and there are no reasons for the light
bending to be negligigle with respect to the other relativistic effects (Doppler boosting,
gravitational redshift, and frame dragging). Another subtle 
issue is the computation of the spectral hardening factor
(here not discussed), which is another weak point of our
approach. Our study has to be taken as a preliminary investigation 
and significant work has still to be done before the continuum 
fitting method can be used to obtain reliable constraints on the Kerr geometry 
around stellar-mass BH candidates.

The content of this paper is as follows. In Sec.~\ref{s-disk}, we
review the basic properties of a geometrically thin and optically
thick accretion disk and how to compute its spectrum. In
Sec.~\ref{s-kerr} and \ref{s-mn}, we present the results of our
calculations, respectively for a Kerr
BH and for a generic object with $q \neq 0$. In 
Sec.~\ref{s-cf}, we show how current observations can be used
to constrain $q$, while in Sec.~\ref{systematics} we discuss the 
possible systematic errors that could affect the continuum fitting technique
and therefore out analysis. Lastly, in~\ref{s-concl} we report our conclusions.
The Manko-Novikov spacetime is reviewed in Appendix~\ref{a-mn}, and the properties of its
ISCO are discussed in Appendix~\ref{a-isco}.

Throughout the paper we use units in which $G_{\rm N}=c=1$, unless stated otherwise.

\section{Thermal spectrum of a thin disk}
\label{s-disk}
The standard model for a geometrically thin and optically
thick accretion disk is due to Novikov and Thorne~\citep{n-t}. In a 
generic stationary, axisymmetric and asymptotically flat spacetime, 
one assumes that the disk is on the equatorial plane, that the disk's 
gas moves on nearly geodesic circular orbits, and that the 
radial heat transport is negligible compared to the energy radiated 
from the disk's surface. From the conservation laws for the 
rest-mass, angular momentum and energy, one can deduce three
basic equations for the time-averaged radial structure of the 
disk~\citep{page-th}. These equations determine the radius-independent time-averaged mass 
accretion rate $\dot{M}$, the time-averaged energy flux $\mathcal{F}(r)$ from the 
surface of the disk (as measured by an observer comoving with the disk's gas) and the time-average torque
$W^r_\phi(r)$:
\be
\dot{M} &=& - 2 \pi \sqrt{-G} \Sigma u^r 
= {\rm const.} \, \\\label{fluxeq}
\mathcal{F}(r) &=& \frac{\dot{M}}{4\pi \sqrt{-G}} f(r) \, , \\
W^r_\phi(r) &=& \frac{\dot{M}}{2\pi\sqrt{-G}} 
\frac{\Omega L_z - E}{\partial_r \Omega} f(r) \, .
\ee
Here $\Sigma$ is the surface density, $u^r$ is the radial 
4-velocity, $G$ is the determinant of the near equatorial plane 
metric in cylindrical coordinates~\footnote{$\sqrt{-G} = \sqrt{\alpha^2 g_{rr} g_{\phi\phi}}$, where 
$\alpha$ is the lapse function. In Kerr spacetime in Boyer-Lindquist 
coordinates, $\sqrt{-G} = r$.}, $E$, $L_z$, and $\Omega$ are 
respectively the conserved specific energy, 
the conserved $z$-component of the specific angular momentum, and
the angular velocity $d\phi/dt$ for equatorial circular geodesics, and $f(r)$ is given by
\be
f(r) = \frac{-\partial_r \Omega}{\left(E 
- \Omega L_z \right)^2} \int_{r_{\rm in}}^{r}
\left(E - \Omega L_z \right) 
\left(\partial_r  L_z \right) \, d\rho \, ,
\ee
where $r_{\rm in}$ is the inner radius of the accretion disk and is 
assumed to be the ISCO radius. More details are given 
in Appendix~\ref{a-isco}.

Since the disk is in thermal equilibrium, the emission is 
blackbody-like and we can define an effective temperature
$T = T(r)$ from the relation $\mathcal{F} = \sigma T^4$,
where $\sigma = 5.67 \times 10^{-5}$~erg~s${-1}$~cm$^2$~K$^{-4}$
is the Stefan-Boltzmann constant. Neglecting the effect of 
light bending, the {\it equivalent isotropic luminosity} is 
\be\label{eq-lum}
L(\nu) = \frac{8\pi h}{c^2} \cos i
\int_{r_{\rm in}}^{r_{\rm out}} \int_{0}^{2\pi}
{\rm g}^3 \frac{ \nu_e^3 \sqrt{- G} dr \, d\phi}{
\exp\left[{h\nu_e}/({kT})\right] - 1} \, ,
\ee
where we have written explicitly the Planck constant
$h$, the speed of light $c$, and the Boltzmann constant $k$. Here,
$i$ is the angle between the distant observer's line of sight and the direction orthogonal to the disk, 
$r_{\rm in}$ and $r_{\rm out}$ are respectively
the inner and outer radius of the disk, while $\nu$ is the radiation frequency 
in the local rest frame of the distant observer 
and $\nu_e$ is the frequency in the local rest frame of the emitter.
These two frequencies are related by the redshift factor
\be
{\rm g} = \frac{\nu}{\nu_e} = 
\frac{k_\mu u^\mu_o}{k_\mu u^\mu_e} \,,
\ee
where $u^\mu_o = (1,0,0,0)$ is the 4-velocity of the observer and $u^\mu_e=(u^t_e,0,0,\Omega u^t_e)$
is the 4-velocity of the emitter. 
Using the normalization condition $g_{\mu\nu}u^\mu_e u^\nu_e = -1$, $u^t_e$
can be obtained to be
\be
u^t_e = \frac{1}{\sqrt{-g_{tt} - 2g_{t\phi}\Omega 
- g_{\phi\phi}\Omega^2}} \, .
\ee
Because the $t$- and $\phi$-component of a photon's canonical
4-momentum are conserved quantities in
any stationary and axisymmetric spacetime, we can
compute the quantity $k_\phi/k_t$ at infinity. The result
is $k_\phi/k_t = r \sin\phi \sin i$ and the redshift factor 
turns out to be
\be
{\rm g} = \frac{\sqrt{-g_{tt} - 2g_{t\phi}
\Omega - g_{\phi\phi}\Omega^2}}{1 + 
\Omega r \sin\phi \sin i} \, .
\ee
With ${\rm g}$, we take into account the special and general
relativistic effects of Doppler boost, gravitational redshift, 
and frame dragging. 

\begin{figure}
\par
\begin{center}
\includegraphics[type=pdf,ext=.pdf,read=.pdf,width=8.0cm]{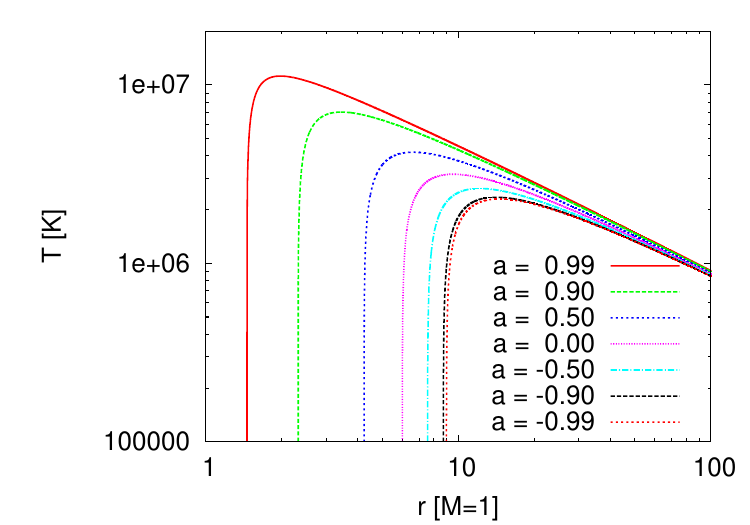}
\includegraphics[type=pdf,ext=.pdf,read=.pdf,width=8.0cm]{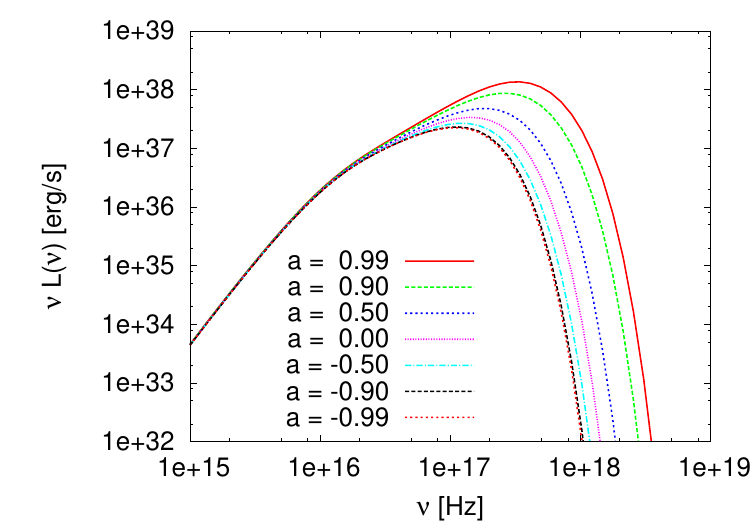}
\end{center}
\par
\caption{Radial profile of the effective temperature (left panel)
and spectrum $\nu L(\nu)$ (right panel) of a thin accretion 
disk in Kerr spacetime for different value of the spin parameter 
$a$. Here we take the mass $M = 10$~$M_\odot$, the mass accretion
rate $\dot{M} = 10^{18}$~g/s, and the inclination angle 
$i = 45^\circ$.}
\label{f-k-spin}
\par
\begin{center}
\includegraphics[type=pdf,ext=.pdf,read=.pdf,width=8.0cm]{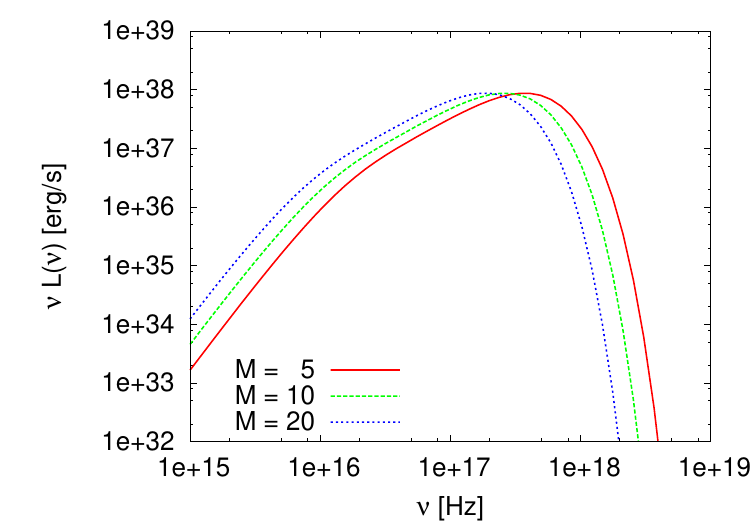}
\includegraphics[type=pdf,ext=.pdf,read=.pdf,width=8.0cm]{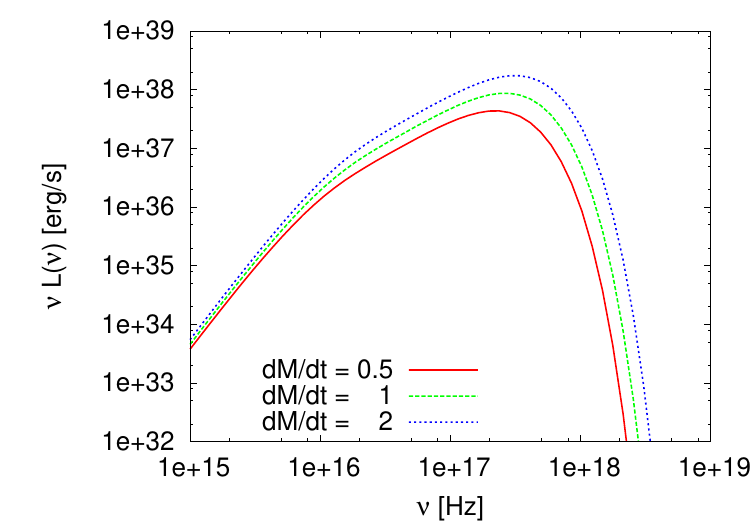}
\end{center}
\par
\caption{Spectrum $\nu L(\nu)$ of a thin accretion 
disk in Kerr spacetime with $a = 0.9$ and an observer inclination
angle $i = 45^\circ$. Left panel: mass $M = 5$, 10, 15~$M_\odot$
and mass accretion rate $\dot{M} = 10^{18}$~g/s. Right panel:
mass $M = 10$~$M_\odot$ and mass accretion rate 
$\dot{M} = 0.5 \times 10^{18}$~g/s, 
$ 10^{18}$~g/s, and $2 \times 10^{18}$~g/s.}
\label{f-k-m}
\end{figure}

\begin{figure}
\par
\begin{center}
\includegraphics[type=pdf,ext=.pdf,read=.pdf,width=8.0cm]{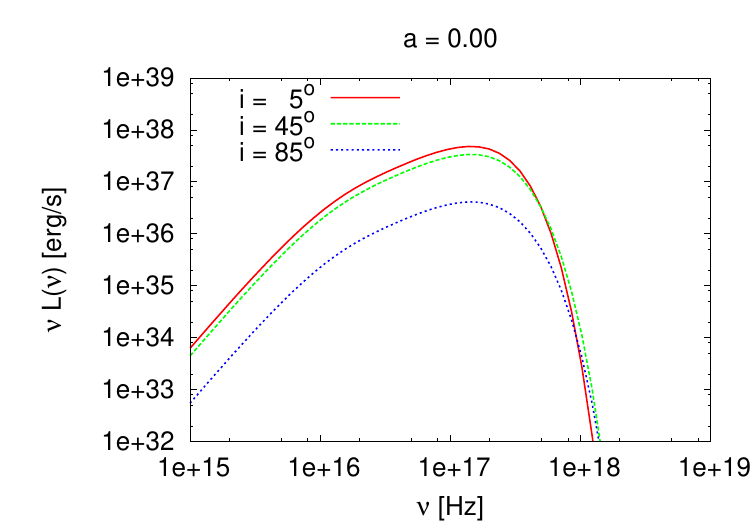}
\includegraphics[type=pdf,ext=.pdf,read=.pdf,width=8.0cm]{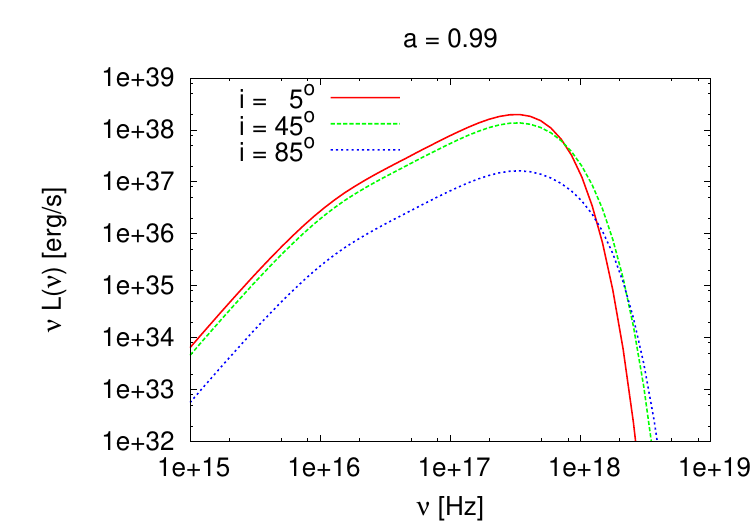}
\end{center}
\par
\caption{Spectrum $\nu L(\nu)$ of a thin accretion 
disk in Kerr spacetime with mass $M = 10$~$M_\odot$ and
mass accretion rate $\dot{M} = 10^{18}$~g/s. Here
we consider different observer inclination angles for
$a = 0.00$ (left panel) and $a = 0.99$ (right panel).}
\label{f-k-i}
\par
\begin{center}
\includegraphics[type=pdf,ext=.pdf,read=.pdf,width=8.0cm]{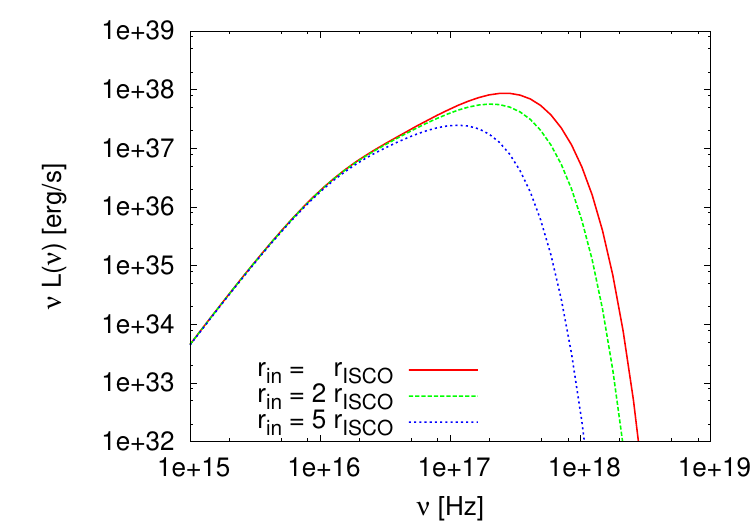}
\includegraphics[type=pdf,ext=.pdf,read=.pdf,width=8.0cm]{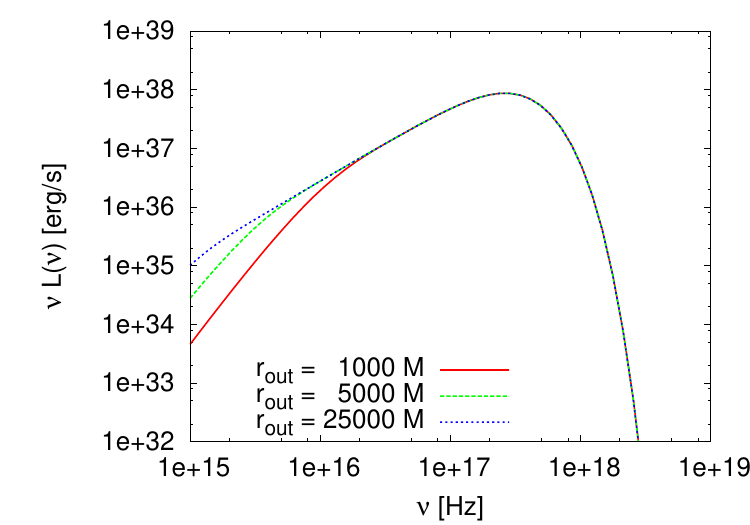}
\end{center}
\par
\caption{Effect of a different inner and outer disk radius.
Left panel: $r_{\rm in} = r_{\rm _{\rm ISCO}}$, $r_{\rm in} = 2 \, r_{\rm _{\rm ISCO}}$, and
$r_{\rm in} = 5 \, r_{\rm _{\rm ISCO}}$. Right panel: $r_{\rm out} = 1000$~$M$, 
$r_{\rm out} = 5000$~$M$, and $r_{\rm out} = 25000$~$M$.}
\label{f-k-rdisk}
\vskip 6mm 
\end{figure}

\section{Spectrum in Kerr spacetimes}
\label{s-kerr}
To begin with, we calculate the thermal spectrum of a geometrically
thin and optically thick accretion disk around a Kerr BH.
Here we have four free parameters determining the 
luminosity~(\ref{eq-lum}):
the mass of the BH $M$, the spin parameter $a$,
the mass accretion rate $\dot{M}$, and the inclination angle of the
disk with respect to the distant observer $i$.
However, usually $M$ and $i$ can be deduced from independent observations (see Section~\ref{s-cf} for an example).

The role of the spin parameter is shown in Fig.~\ref{f-k-spin}, 
where we assume $M = 10$~$M_\odot$, $\dot{M} = 10^{18}$~g/s, 
and $i = 45^\circ$. In the left panel, we present the radial 
profile of the effective temperature and, in the right panel, 
the observed spectrum $\nu L(\nu)$. For $a < 0$, we mean that 
the disk is counterrotating. Since we assume that 
$r_{\rm in} = r_{\rm _{\rm ISCO}}$, the spin parameter determines the inner radius 
of the disk: as $a$ increases, $r_{\rm in}$ decreases and we find
warmer matter at smaller radii. At larger radii the effective
temperature is essentially independent of the spin parameter.
Therefore, a higher spin parameter moves the peak of $\nu L(\nu)$ to
higher frequency and to higher values.

Changing the BH mass while keeping $\dot{M}$ constant\footnote{While this assumption
is useful to single out the effect of a change in $M$, if a BH accretes at the Eddington rate one has $\dot{M}\propto M$.} 
has two effects. For larger
masses, the effective temperature 
decreases ($T \propto M^{-1/2}$) and therefore the peak 
of spectrum moves to lower frequency. At the same time, the size
of the disk increases, increasing the total luminosity. In the
left panel of Fig.~\ref{f-k-m} we show the cases $M = 5$, 10, 
15~$M_\odot$ for $a = 0.9$. The role of the mass accretion rate
is shown in the right panel of Fig.~\ref{f-k-m}. It is clear that a change in $\dot{M}$ only changes the 
effective temperature.

The viewing angle $i$ determines the effective disk surface 
seen by the distant observer and the correction due to the
Doppler boosting (for $i = 0^\circ$ there is no Doppler 
boosting)\footnote{We remind the reader that here we neglect 
the effect of light bending, whose contribution would also
depend on the viewing angle.}.
In Fig.~\ref{f-k-i} we show the cases $i = 5^\circ$, 45$^\circ$
and 85$^\circ$ for $a = 0$ (left panel) and $a = 0.99$
(right panel).

\begin{figure}
\par
\begin{center}
\includegraphics[type=pdf,ext=.pdf,read=.pdf,width=8.0cm]{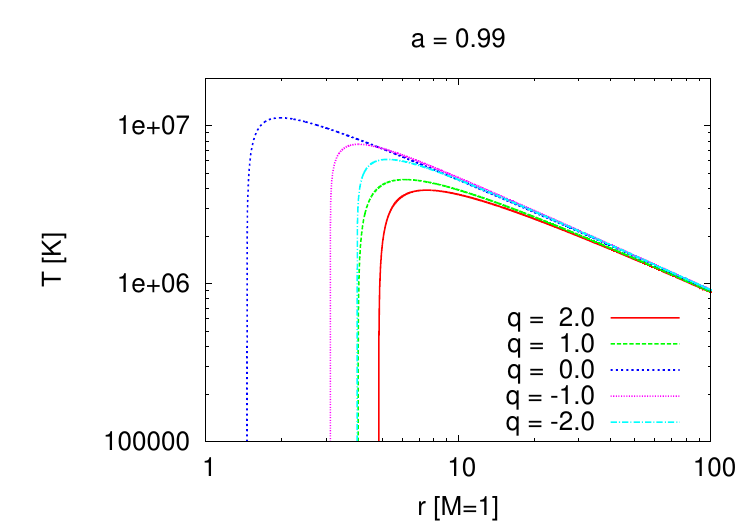}
\includegraphics[type=pdf,ext=.pdf,read=.pdf,width=8.0cm]{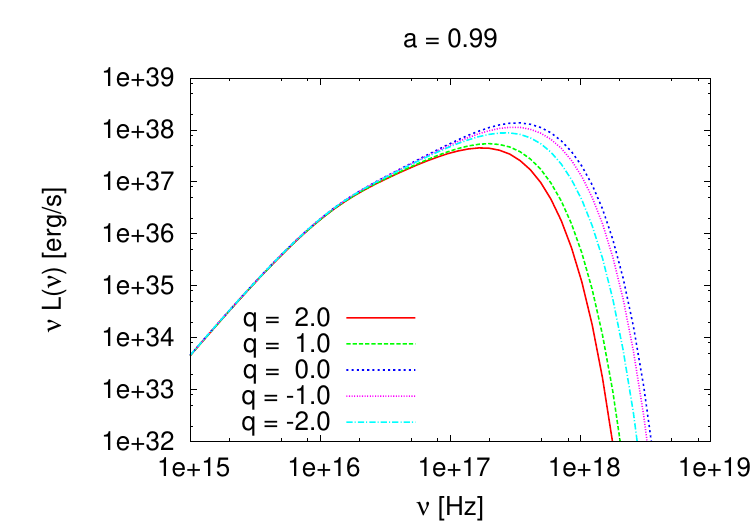} \\
\includegraphics[type=pdf,ext=.pdf,read=.pdf,width=8.0cm]{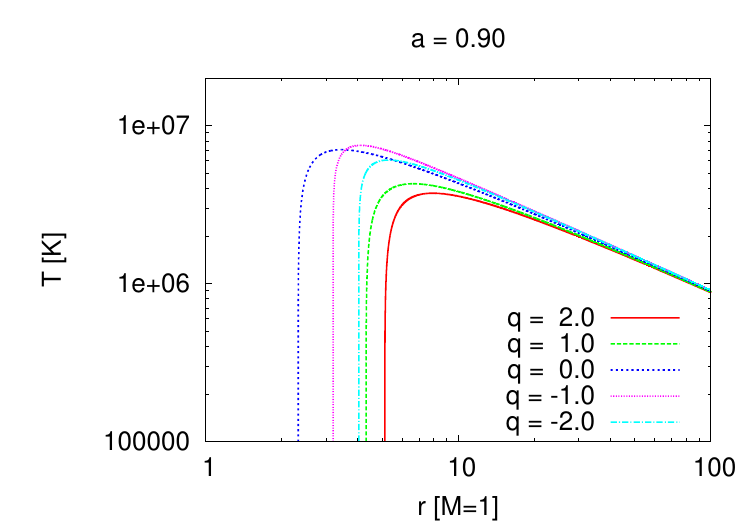}
\includegraphics[type=pdf,ext=.pdf,read=.pdf,width=8.0cm]{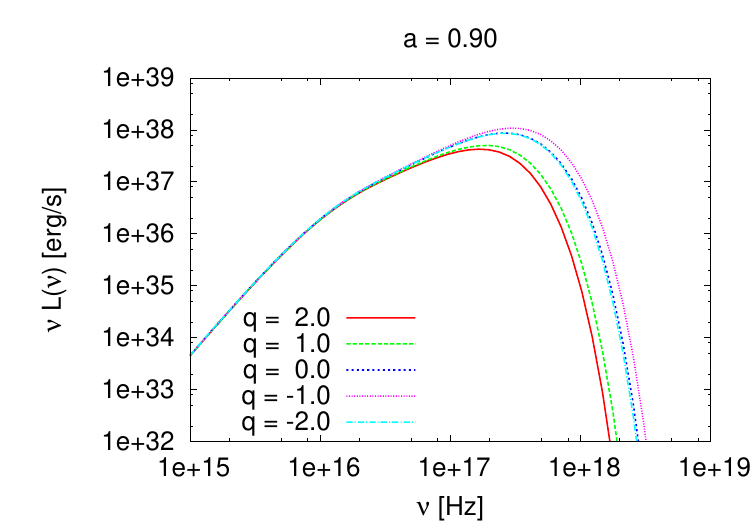} \\
\end{center}
\par
\caption{Radial profile of the effective temperature (left panels)
and spectrum $\nu L(\nu)$ (right panels) of a thin accretion 
disk in Manko-Novikov spacetime for spin parameter $a = 0.99$ 
(top panels) and $a = 0.90$ (bottom panels) and for different 
values of the anomalous quadrupole moment $q$. Here we take the 
mass $M = 10$~$M_\odot$, the mass accretion rate 
$\dot{M} = 10^{18}$~g/s, and the inclination angle $i = 45^\circ$.}
\label{f-mna}
\vskip8mm
\end{figure}

\begin{figure}
\par
\begin{center}
\includegraphics[type=pdf,ext=.pdf,read=.pdf,width=8.0cm]{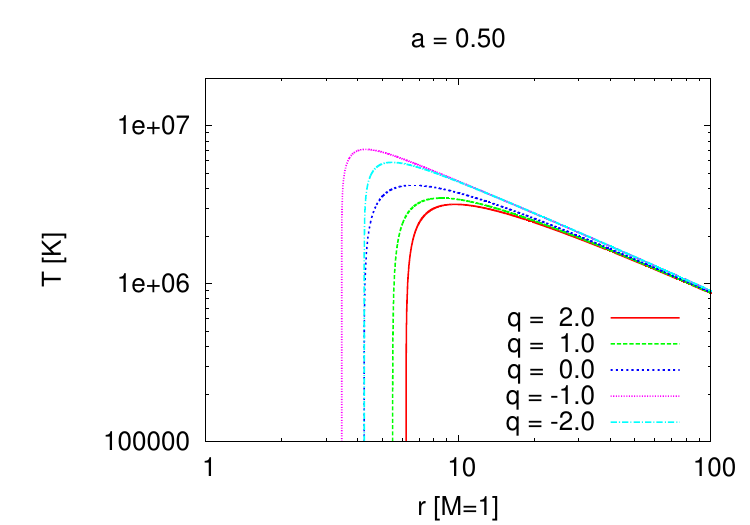}
\includegraphics[type=pdf,ext=.pdf,read=.pdf,width=8.0cm]{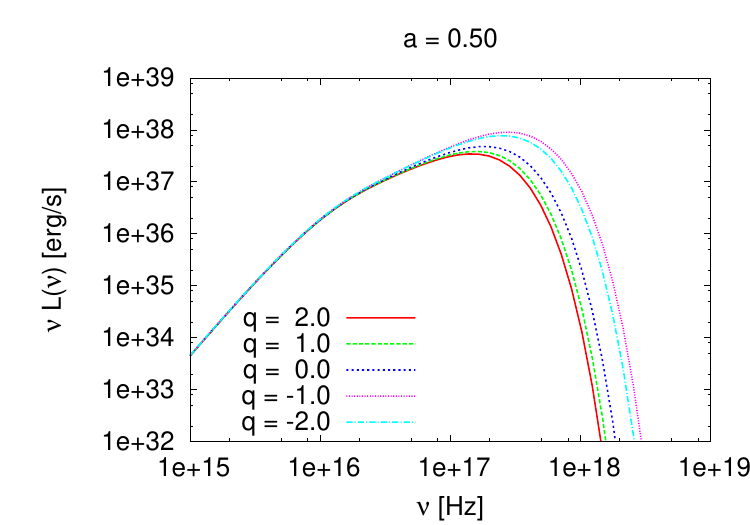} \\
\includegraphics[type=pdf,ext=.pdf,read=.pdf,width=8.0cm]{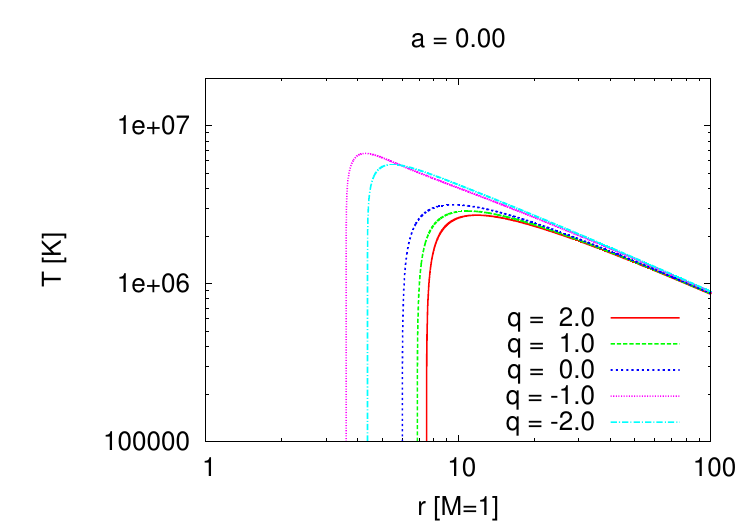}
\includegraphics[type=pdf,ext=.pdf,read=.pdf,width=8.0cm]{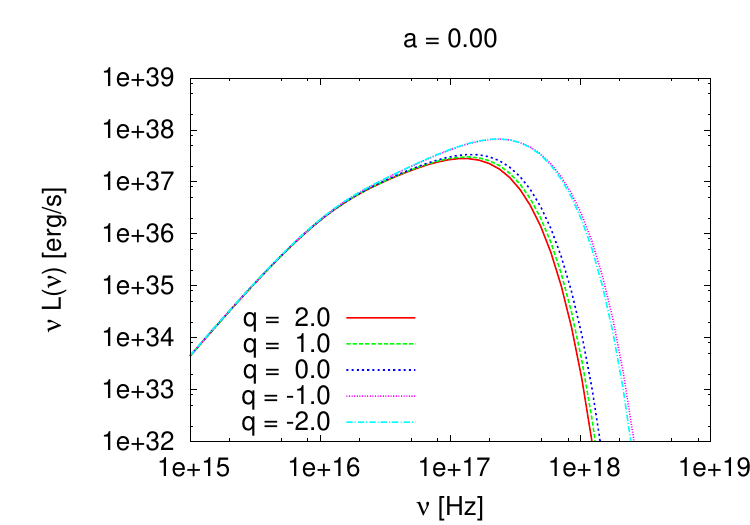} \\
\end{center}
\par
\caption{As in Fig.~\ref{f-mna}, for $a = 0.50$ (top panels)
and $a = 0.00$ (bottom panels).}
\label{f-mnb}
\vskip8mm
\end{figure}

\begin{figure}
\par
\begin{center}
\includegraphics[type=pdf,ext=.pdf,read=.pdf,width=8.0cm]{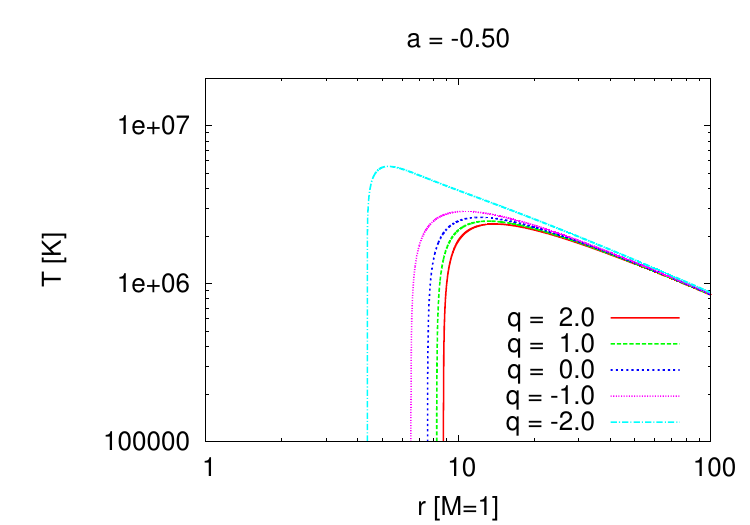}
\includegraphics[type=pdf,ext=.pdf,read=.pdf,width=8.0cm]{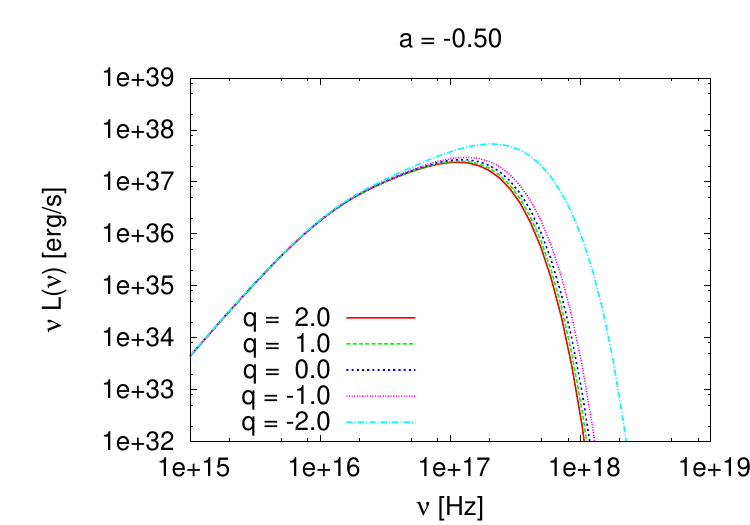} \\
\includegraphics[type=pdf,ext=.pdf,read=.pdf,width=8.0cm]{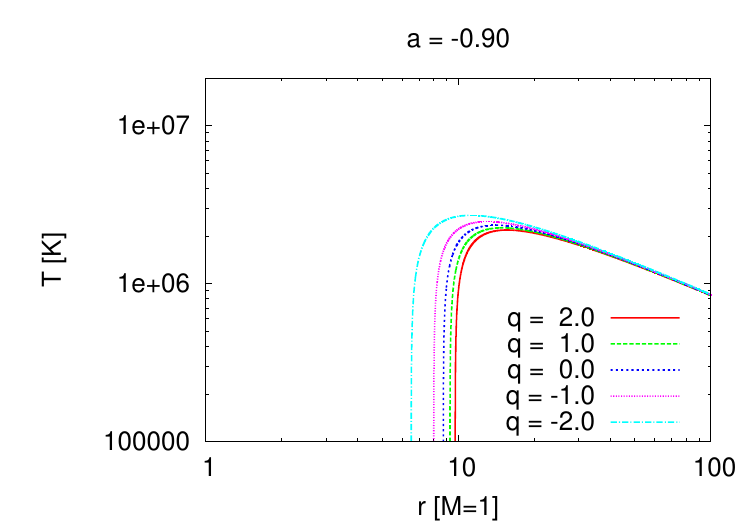}
\includegraphics[type=pdf,ext=.pdf,read=.pdf,width=8.0cm]{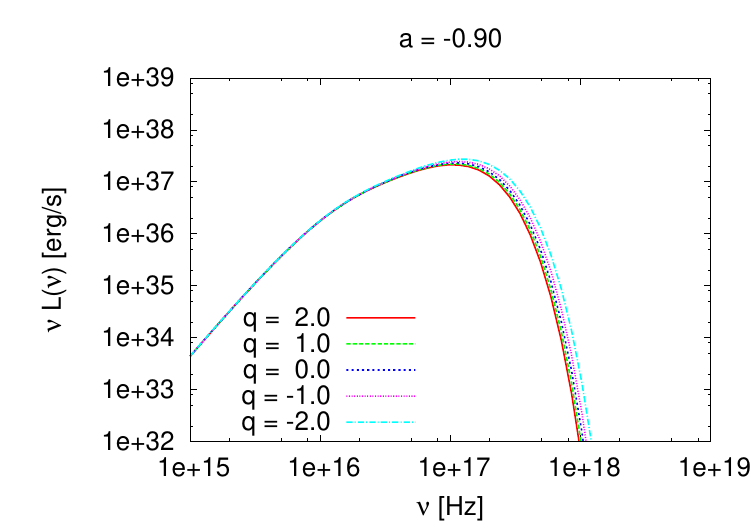}
\end{center}
\par
\caption{As in Fig.~\ref{f-mna}, for $a = 0.50$ (top panels) and 
$a = 0.90$ (bottom panels) in the case of counterrotating disks.}
\label{f-mnc}
\end{figure}

Lastly, we show the effect of $r_{\rm in}$ and $r_{\rm out}$ on the shape
of the spectrum. So far we have adopted the standard
assumption that the inner radius of the disk is at the ISCO and
we have chosen the outer radius $r_{\rm out} = 10^3$~$M$. However,
if $r_{\rm in}$ were larger than the radius of the ISCO, it would
affect the high frequency part of the spectrum, mimicking a
lower spin parameter, see Fig.~\ref{f-k-rdisk}. On the other
hand, assuming a larger outer radius of the disk, the
spectrum moves the cut-off at lower frequencies,
with no changes at higher frequencies.

\section{Spectrum in Manko-Novikov spacetimes}
 \label{s-mn}
Let us now consider the more general case in which the compact
object is not a Kerr BH. The gravitational field
around a generic compact body can be described by the 
Manko-Novikov metric, which is a stationary, axisymmetric and
asymptotically flat exact solution of the vacuum Einstein 
equation and has an infinite number of free parameters. The
structure of the spacetime presents strong similarities with
the $\delta = 2$ Tomimatsu-Sato spacetime~\citep{t-s, kodama}. Here
we restrict our attention to a subclass of the Manko-Novikov
solution, where the compact object is determined 
by its mass $M$, its spin parameter $a = J/M^2$, and the 
anomalous quadrupole moment $q$ which regulates the deviations of the spacetime
from the Kerr geometry:
\be\label{qdef}
q = - \frac{Q - Q_{\rm Kerr}}{M^3} \, ,
\ee 
where $Q$ and $Q_{\rm Kerr} = - a^2 M^3$ are respectively 
the quadrupole moment of the object and that of a Kerr 
BH. In this work, we consider only spin parameters
$|a| \le 1$, which is the allowed range in the standard Manko-Novikov
solution\footnote{In the Kerr case, $|a| \le 1$ is the
condition for the existence of an event horizon. However,
in the more general case of a compact object made of some
kind of exotic matter, the maximum value of $|a|$ may in principle be
larger than 1~\citep{cb0,cb1,cb2,cb3,cb4}.
As in the case of the Tomimatsu-Sato family, also the Manko-Novikov solution
can probably be  extended to describe objects with
$|a|>1$~\citep{moreno}. However, if such fast-rotating objects
are very compact, they are most likely unstable, 
at least for small values of $q$, due to the ergoregion instability~\citep{pani}.}. As for the value of the anomalous 
quadrupole moment  measuring the deviation from the Kerr 
metric: for $q > 0$ the object is more oblate than a Kerr
BH; for $q < 0$ it is more prolate; for $q = 0$
the Manko-Novikov metric reduces \textit{exactly} to the Kerr metric.
Thanks to this property, this subclass of the Manko-Novikov solution, for which we give complete expressions in Appendix~\ref{a-mn},
is a perfect tool to set up a null experiment~\citep{scott_procs} to test the validity of the Kerr
metric and of the no-hair theorem: any experiment pointing at a significantly non-zero value 
for $q$ would imply that the compact object under consideration is not a BH as described by General Relativity. 
This use of the Manko-Novikov metric has been put forward in 
gravitational-wave astrophysics, namely in \citet{gair}. See~\citet{bumpy1,bumpy2,kostas}
for other metrics which reduce exactly to the Kerr metric when the equivalent of
our anomalous quadrupole parameter $q$ is set to $0$, and which can 
therefore be used to perform null experiments testing the Kerr geometry.\footnote{
The ``bumpy BHs'' of
\citet{bumpy1,bumpy2} seem related to the Manko-Novikov general solution because
they too allow an arbitrary multipolar structure for the non-Kerr
spacetime. The ``quasi-Kerr'' metric of \citet{kostas}, instead,
is an (approximate) solution of the vacuum Einstein equations only for small values of the spacetime's spin,
and allows only quadrupolar deformations. For these
reasons, bumpy BHs or Manko-Novikov spacetimes are preferable options
to test general deviations from the Kerr metric.}

In Figs.~\ref{f-mna}, \ref{f-mnb}, and \ref{f-mnc},
we show the radial profile of the thin accretion disk's
effective temperature and the spectrum $\nu L(\nu)$
for a few values of $a$ and $q$.
We still assume $M = 10$~$M_\odot$, $\dot{M} = 10^{18}$~g/s,
and $i = 45^\circ$. For a given spin parameter, the value of $q$
determines the radius of the ISCO -- see also Appendix~\ref{a-isco}.
Since the temperature of the disk is higher at small radii,
a non-zero $q$ produces corrections in the high frequency
region of the spectrum, while at low frequencies there are
no changes. The effect is quite small for slow-rotating objects
or counterrotating disks, while it becomes relevant, and actually
non-negligible, for fast-rotating bodies and corotating disks. 
There are two
reasons for this: a small deviation from $q \neq 0$ produces a larger 
variation in the radius of the ISCO for higher spin parameters --
see Fig.~\ref{f-isco} in Appendix~\ref{a-isco} --  and, because the
ISCO is closer to the compact object as $a$ approaches 1, the 
spectrum of the disk is more sensitive to small deviations in 
the multipole moment expansion. This is the contrary of what
happens in the Tomimatus-Sato spacetimes, where for $|a| \rar 1$
all the solutions reduce to an extreme Kerr BH and
thus deviations from the Kerr metric are more relevant for low
spin parameters~\citep{naoki}.

\section{Observational constraints: the case of M33~X-7}
\label{s-cf}
As an example of how accretion-disk thermal spectra can already
put significant constraints on the deviation $q$ of the quadrupole 
moment of BH candidates from that of a Kerr BH (\textit{cf.} Eq.~(\ref{qdef})),
we consider the case of M33~X-7. This object is an eclipsing X-ray binary 
consisting of a BH candidate accreting from a companion star~\citep{M33X7},
and its orbital parameters and its distance are measured with 
the highest accuracy among all known BH binaries~\citep{M33X7accuracy} (see Table \ref{pubres}). 
In particular, the BH candidate's mass is measured
to be $M=15.65\pm1.45 M_\odot$, while the disk's inclination is $i=74.6^\circ\pm 1^\circ$
and the distance is $d=840\pm 20$ kpc~\citep{M33X7accuracy}. 

The accurate knowledge of $M$, $i$ and $d$ allows 
the continuum fitting method~\citep{zhang} to extract reliable 
information on the spin of the BH candidate. By essentially 
fitting the Chandra and XMM-Newton spectra of M33~X-7
with a relativistic accretion disk model depending on the spin $a$, 
the Eddington ratio $\ell=L_{\rm bol}/L_{\rm Edd}$ (where $L_{\rm bol}$ 
and $L_{\rm Edd}=1.2572\times 10^{38} (M/M_\odot)$ erg/s are the bolometric and Eddington luminosities) 
and the hydrogen column density $N_{\rm H}$, \citet{m33x7,m33x7e} measured the spin to be $a=0.84\pm0.05$.\footnote{We 
warn the reader that this is the revised value reported in \citet{m33x7e}, which corrects a bug in the analysis of \citet{m33x7}.} 
The Eddington ratio is instead $\ell=0.0989\pm 0.0073$ 
(\textit{cf.} Table I of \citet{m33x7} and \citet{m33x7e}).
We notice that
the errors on $a$ and $\ell$ include also the (propagated) effect of the uncertainties on
$M$, $d$ and $i$~\citep{m33x7,m33x7e}.

Our simple disk model depends on three parameters, $a$, $q$ and 
the Eddington ratio $\ell=L_{\rm bol}/L_{\rm Edd}$. The latter can be
rexpressed as $\ell=\dot{M}/\dot{M}_{\rm Edd}(a,q)$, where we define 
the Eddington accretion rate as 
\begin{equation}
\dot{M}_{\rm Edd}(a,q)=\frac{L_{\rm Edd}}{\eta(a,q) c^2}\,,
\end{equation}
$\eta=1-E_{\rm _{\rm{ISCO}}}(r_{\rm _{\rm{ISCO}}})$ being 
the efficiency of the conversion between rest-mass and electromagnetic energy~\citep{GRbook}.
Ideally we would then have to fit the observed spectrum of M33~X-7 with this 3-parameter model. 
However, because of the difficulties 
and subtleties of analyzing the real Chandra and XMM-Newton spectra and because
 of the simplified nature of our disk model, we resorted 
to a simpler approach. While a thorough analysis of the real data 
will be needed to determine the precises constraints on $q$, our simplified
treatment will show that such an analysis is definitely worth being done as 
it would permit ruling out entire regions of the $(a,q)$ plane.

In particular, instead of comparing our disk model with the raw data, 
we compare it to the spectrum of a thin disk with
$\ell^\star=0.0989$ and inclination $i=74.6^\circ$ around a Kerr BH with 
spin $a^{\star}=0.84$ and mass $M=15.65M_\odot$ (these are the values measured
for M33~X-7).
While meaningful and reliable constraints on the parameter
$a$ and $q$ can only be obtained by fitting the original X-ray data,
we use here this simplified approach because ours is a preliminary investigation and our results are only meant
as a qualitative guide for future more rigorous studies.
The spectrum is calculated with the standard Novikov-Thorne model reviewed in section \ref{s-kerr}.
The observational errors on the ``measured'' spectrum are then mimicked by using the 
estimated final errors on the spin ($\delta a=0.05$) and Eddington ratio ($\delta \ell=0.0073$).
Because the Eddington ratio regulates the bolometric luminosity (\textit{i.e.,} the normalization of the spectrum)
one has $L^{\rm Kerr}(\nu,a^{\star},\ell^{\star})<L^{\rm Kerr}(\nu,a^{\star}\pm\delta a,\ell^{\star}+\delta\ell)$
and $L^{\rm Kerr}(\nu,a^{\star},\ell^{\star})>L^{\rm Kerr}(\nu,a^{\star}\pm\delta a,\ell^{\star}-\delta\ell)$. It therefore makes
sense to define the error as 
\begin{gather}\label{err}
\sigma(\nu)=\frac{\max(\nu L^{\rm Kerr}(\nu,a^{\star}\pm\delta a,\ell^{\star}+\delta\ell))-
\min(\nu L^{\rm Kerr}(\nu,a^{\star}\pm\delta a,\ell^{\star}-\delta\ell))}{2}\,.
\end{gather}
To determine the values of $a$, $q$ and $\ell$ giving the best fit, one would then have to minimize the reduced $\chi^2$, which we define 
as
\be
\chi_{\rm red}^2 =\frac{\chi^2}{N}=\frac{1}{N}\sum^{i=N}_{i=1} 
\left(\frac{\nu_i L^{\rm MN}(\nu_i,a,q,\ell) - \nu_i L^{\rm Kerr}(\nu_i,a^\star,\ell^\star)}{\sigma(\nu_i)}\right)^2 
\ee
where the summation is performed over $N$ sampling frequencies $\nu_i$
and where $L^{\rm MN}$ and $ L^{\rm Kerr}$ are calculated as explained in Sections~\ref{s-mn} and~\ref{s-kerr} respectively. 
To simplify the analysis, here we assume that the Eddigton ratio is fixed to the 
measured value $\ell^\star$, and therefore seek to minimize
\be\label{chi2}
\chi_{\rm red}^2 =\frac{\chi^2}{N}=\frac{1}{N}\sum^{i=N}_{i=1} 
\left(\frac{\nu_i L^{\rm MN}(\nu_i,a,q,\ell^\star) -\nu_i L^{\rm Kerr}(\nu_i,a^\star,\ell^\star)}{\sigma(\nu_i)}\right)^2 \,.
\ee
We stress that this simplified approach makes sense because in principle the Eddington 
ratio can be determined independently, by integrating the luminosity
over all frequencies.

Since the spectrum of M33~X-7 is well fit  with an accretion
disk around a Kerr BH~\citep{m33x7,m33x7e}, one may wonder whether a
fit with an additional parameter is statistically justified. As already stressed, our
viewpoint is that X-ray continuum spectra can be used to constrain 
the value of the anomalous quadrupole moment, precisely because 
of the data's small error bars. This is, in other words, a null
experiment, {\it i.e.} one in which we seek to measure a quantity 
which we expect to be zero. The measurement is therefore not one 
of the quantity itself, but rather a measurement of its ``error 
bars'' around its expected zero value.

\begin{table}
\begin{center}
\begin{tabular}{r c r c r c l}
\hline \\
Binary System & \hspace{.5cm} & $M/M_\odot$ & \hspace{.5cm} & 
$a$ & \hspace{.5cm} & Reference \\ \\
\hline 
4U~1543-47 & & $9.4 \pm 1.0$ & & $0.75-0.85$ & & 
\citet{l113} \\
GRO~J1655-40 & & $6.30 \pm 0.27$ & & $0.65-0.75$ & & 
\citet{l113} \\ 
GRS~1915+105 & & $14.0 \pm 4.4$ & & $> 0.98$ & & 
\citet{1915} \\ 
LMC~X-3 & & $5 - 11$ & & $< 0.26$ & &
\citet{done} \\ 
M33~X-7 & & $15.65 \pm 1.45$ & & $0.84 \pm 0.05$ & & 
\citet{m33x7,m33x7e} \\ 
LMC~X-1 & & $10.91 \pm 1.41$ & & $0.92^{+0.05}_{-0.07}$ & & 
\citet{gou} \\
XTE~J1550-564 & & $9.10 \pm 0.61$ & & $0.34^{+0.20}_{-0.28}$ & & 
\citet{steiner-b} \\
\hline
\end{tabular}
\end{center}
\caption{Published spin measurements of stellar-mass BH
candidates with the continuum fitting method.}
\label{pubres}
\end{table}

\begin{figure}
\par
\begin{center}
\includegraphics[type=pdf,ext=.pdf,read=.pdf,width=7.75cm]{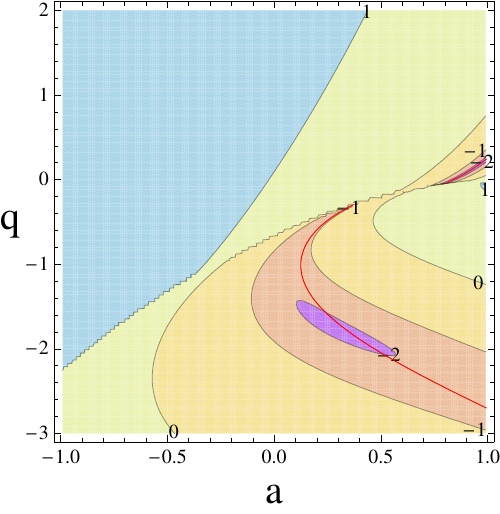}\hskip 1cm 
\includegraphics[type=pdf,ext=.pdf,read=.pdf,width=7.75cm]{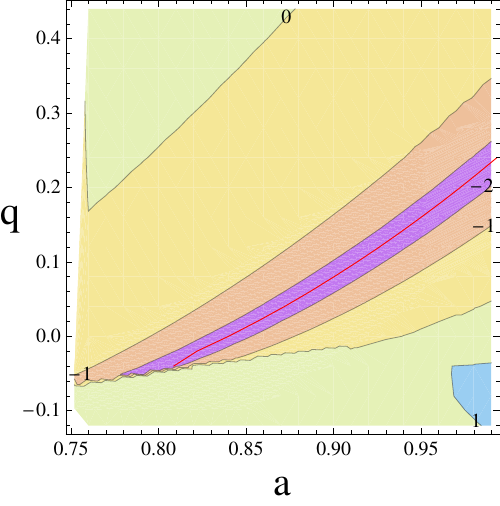} 
\end{center}
\par
\caption{\label{whole} $\log_{10}(\chi^2_{\rm red})$, as defined by Eq.~(\ref{chi2}), for the comparison between
the spectrum of M33~X-7 and that of a thin accretion disk in a Manko-Novikov spacetime with 
spin $a$ and quadrupole parameter $q$ (as defined in Eq.~(\ref{qdef})). 
Instead of fitting the original X-ray data, we use a simplified model for
the spectrum of  M33~X-7 (see text for details), and therefore the constraints 
$a$ and
$q$ only have a qualitative meaning.
The viable regions are those with $\log_{10}(\chi^2_{\rm red})<0$, hence
ruling out roughly half of the $(a,q)$ plane. 
Our naive assumption (\ref{err}) for the errors probably overstimates the real observational uncertainties (see text for details).
If the error (\ref{err}) were too large by a factor $\sqrt{10}\approx3.16$ (10), $\chi_{\rm red}^2$ 
would decrease by a factor 10 (100), effectively
restricting the allowed $(a,q)$ to the regions with $\log_{10}(\chi_{\rm red}^2)<-1$ ($\log_{10}(\chi_{\rm red}^2)<-2$).
The red line denotes the $(a,q)$ for which the efficiency $\eta(a,q)$ equals that of M33~X-7 (see text for a physical interpretation).
}
\end{figure}

In Fig.~\ref{whole} we show the contour plots of $\log_{10}(\chi_{\rm red}^2)$, as a function of $a$ and $q$ and as given by Eq.~(\ref{chi2}),
in which we choose $N=41$ sampling frequencies $\nu_i$ equally spaced, in logarithmic scale, from $10^{15}$ to $1.5\times10^{18}$ Hz. Also, to calculate 
$L^{\rm MN}$ and $L^{\rm Kerr}$ we assume $r_{\rm out}=10^3 M$, which allows us to significantly reduce the computational time
needed to produce the spectra with respect to
a larger outer radius.
The regions of parameter space which are viable present $\log_{10}(\chi_{\rm red}^2)<0$. 
As expected,  $\chi_{\rm red}^2$ presents a minimum around $a^\star=0.84$, $q=0$ (due to Eq.~(\ref{chi2}), $\chi_{\rm red}^2$
is exactly $0$ there), but also the surrounding ``valley'' 
is in agreement with the data (see the right panel of Fig.~\ref{whole}). 
Moreover, a larger ``valley'' (featuring a central ``basin'') with $\chi_{\rm red}^2<1$ exists for $q\lesssim-0.3$ and $a\gtrsim-0.5$, 
separated from the first one by a saddle.

The physical interpretation of these two allowed ``valleys'' is quite straightforward. They stretch across
the red line in Fig.~\ref{whole}, which corresponds the $(a,q)$ for which $\eta(a,q)=\eta(a=a^\star,q=0)$, 
where $\eta(a=a^\star,q=0)$ is the efficiency of the Kerr model that we use to mimick the 
data for M33 X-7.\footnote{The redline disappears for $-0.29\lesssim q\lesssim-0.05$ because of the discontinous dependence of the ISCO on
$(a,q)$ (see Appendix \ref{a-isco}).} 
This fact is easy to understand. In our analysis we assume 
that the bolometric luminosity, $L_{\rm bol}=\eta(a,q) \dot{M} c^2$ is fixed and given by $\ell^\star L_{\rm edd}=\eta(a=a^\star,q=0) \dot{M}^\star c^2$,
where $\dot{M}^\star$ denotes the accretion rate of M33 X-7. The accretion rate $\dot{M}$ is constrained to be close to $\dot{M}^\star$
in order for the Manko-Novikov spectrum to reproduce that of M33 X-7 at low frequencies. This is because $\dot{M}$ basically regulates
the slope of the spectrum at low frequencies: from Eq.~(\ref{eq-lum}) one 
gets $L(\nu)\sim T$ at small frequencies, but $T\propto \dot{M}^{1/4}$ because of Eq.~(\ref{fluxeq}) and the blackbody assumption.
Therefore, if $\dot{M}\sim\dot{M}^\star$ one obtains that it must be $\eta(a,q)\sim\eta(a=a^\star,q=0)$.

We stress that we have determined these two allowed regions under the conservative assumption (\ref{err}) for the error $\sigma$.
In Eq. (\ref{err}) we basically assumed that the errors determined by \citet{m33x7,m33x7e} for $\ell$ and $a$ were uncorrelated, 
which could result in an estimate slightly larger than the real observational errors. This is hinted at also by
Fig.~\ref{whole}. If one assumes $q=0$ (\textit{i.e.,} if one adopts the Kerr-BH hypothesis) Fig.~\ref{whole} shows that the allowed spins
would be $0.65\lesssim a\lesssim0.95$, whereas \citet{m33x7,m33x7e} find $a^\star=0.84\pm0.05$. 
If our naive assumption overstimated the real observational errors by a factor $\sqrt{10}\approx3.16$, $\chi_{\rm red}^2$ 
would decrease by a factor 10, effectively
restricting the allowed $(a,q)$ to the regions of Fig.~\ref{whole} where $\log_{10}(\chi_{\rm red}^2)<-1$. One can see that
for $q=0$,  $\log_{10}(\chi_{\rm red}^2)<-1$ would indeed give $0.79\lesssim a\lesssim0.88$, similar to the interval identified by 
\citet{m33x7,m33x7e}.
Likewise, an observational error
10 times smaller than Eq.~(\ref{err}) would constrain the viable models to the regions with $\log_{10}(\chi_{\rm red}^2)<-2$.

Even with our conservative assumption for the errors, however,  Fig.~\ref{whole} shows that more than half of the $(a,q)$ plane is ruled out.
Nevertheless, if systematic errors for M33~X-7 were larger than assumed in \citet{m33x7,m33x7e}, the constraints might be considerably 
weaker. For example, if the errors were $\sqrt{10}\approx3.16$ larger than our assumption, only the region with
$\log_{10}(\chi_{\rm red}^2)>1$ would be ruled out. We discuss possible sources of systematic error in the next section. We 
stress, however, that the presence of significant systematics would not only jeopardize our test of the no-hair theorem, but would 
represent a very serious problem also for the spin measurements with the continuum fitting technique, even if the Kerr-BH hypothesis is adopted.
(This can be understood by looking at Fig.~\ref{whole} for $q=0$: as can be seen, the allowed interval for $a$ grows rapidly if the error increases.)

\section{Possible sources of systematic errors}
\label{systematics}

The continuum fitting method is a very promising technique 
for probing the space-time of stellar-mass BH candidates. 
Nevertheless, it is important to keep in mind that there are sources of 
of systematic errors that still need to be understood in order 
 to obtain robust estimates of the spin parameter 
(if one assumes the Kerr-BH hypothesis) or constraints on the anomalous
quadrupole moment with this method.

The main source of uncertainty is the estimate of the hardening factor,
sometimes called color factor, $f_{col}$. Because in the inner part of 
the disk the temperature exceeds $10^6$~K, non-thermal processes are
non-negligible and the spectrum observed by a distant 
observer is not the blackbody-like spectrum computed from the disk's effective
temperature $T$. The hardening factor takes this effect 
into account, by replacing $T$ with 
the color temperature $T_{col} = f_{col} T$, and its 
typical values are in the range $f_{col}= 1.5 - 2.0$. The
computation of the hardening factor requires a reliable model of the disk
atmosphere and its importance has been already 
stressed in~\citet{li05}. Significant progresses to address
this issue have been done in \citet{davis} and in \citet{davis2}.

The continuum fitting technique also assumes that the spin of the 
compact object is perpendicular to the inner part of the accretion 
disk to within a few degrees. For stellar-mass BH candidates in 
X-ray binary systems, we expect this to be true, on the basis of binary population
synthesis~\citep{bps}. While the Bardeen-Petterson effect~\citep{b-p} 
may also be responsible for this effect, for young objects the 
timescale necessary to align the central part of the disk turns out to 
be too long. However, there are also observational data~\citep{tilt1} 
and theoretical arguments~\citep{tilt2} suggesting that tilted disks 
may be possible. This assumption will be
checked by future X-ray polarimetry 
observations~\citep{li09,schnittman09,schnittman10}, such as the 
GEMS mission scheduled for 2014.

In our current analysis, we have also neglected the effect of light
bending, because this is just a preliminary study to determine whether the
continuum fitting method can conceivably be used to constrain deviations from the
Kerr metric. While the effect of light bending must be taken into
account in a complete analysis of the observational data, it has been quite
commonly neglected in similar preliminary studies appeared in 
the literature~\citep{harko_gs,harko_ws,harko_hl,harko_cs,harko_ns}.

\section{Conclusions}
\label{s-concl}
If current astrophysical BH candidates are Kerr BHs,
their spacetime should be completely specified by two
parameters, namely their mass $M$ and spin $J$. This can be tested by measuring at least three multipole
moments of the BH candidate. While there are several proposals to obtain
such a measurement with future experiments, in this paper we have shown that current X-ray observations
of stellar-mass BH candidates in binary systems can
already be used to constrain possible deviations
from the Kerr metric.

We have computed the thermal spectrum of a geometrically thin and 
optically thick accretion disk around a compact object with mass
$M$, spin parameter $|a| \le 1$ and arbitrary anomalous quadrupole 
moment $q$. For $q = 0$, we recover the Kerr metric. The exact 
value of $q$ determines the inner radius of the disk, changing the 
high frequency region of the spectrum. The effect is small for low 
spin parameters or for counterrotating disk, but it becomes 
important for higher values of $a$. In general, the sole analysis 
of the disk spectrum cannot completely determine $q$,
because the spectrum is degenerate in $a$ and $q$ and 
therefore one would need an independent measure of $a$. However,
for very fast-rotating Kerr BHs the ISCO radius and therefore the disk's inner radius
becomes very small. Any deviation from $q=0$ (\textit{i.e.,} from the Kerr solution) makes the
inner radius grow quickly. Since current observations suggest that
the inner radius of the accretion disk of some stellar-mass BH 
candidates is close to the gravitational radius $R_g=G M/c^2$, 
one can constrain the anomalous
quadrupole moment of these objects very efficiently.

In this paper we have considered a specific example, the stellar-mass 
BH candidate M33~X-7, whose estimated spin is $a = 0.84 \pm 0.05$
if one assumes it is a Kerr BH~\citep{m33x7,m33x7e}.
Since stronger constraints on $q$ can be
obtained from objects with higher $a$, we could have considered GRS~$1915+105$,
whose spin parameter has been estimated to be larger than 0.98
in~\citet{1915} under the Kerr-BH assumption. However, the measurements of the distance,  mass and 
viewing angle of 
M33~X-7 are more reliable, thus making this object more suitable to
obtain preliminary constraints on the 
$a-q$ plane.

To move our analysis beyond the simplified and preliminary stage
we achieved in this paper, it is of paramount importance to properly understand the
systematic errors that might affect the continuum fitting method, and which could in principle
blur the difference between the spectra of Kerr BHs and those of other objects, and affect
the measurements of the spin even if one adopts the Kerr BH hypothesis.
Moreover, we will have to amend our disk model by 
including the following ingredients:
\begin{enumerate}
\item {\it The effect of light bending}. 
A rigorous computation of the spectrum requires to trace the 
light rays from the surface of the accretion disk to the distant 
observer in the background metric. The effect of light bending is
presumably no less important than the other relativistic effects
and further alters the observed spectrum.
\item {\it The spectral hardening factor}. In the inner part of the
accretion disk, the temperature is high and non-blackbody effects
cannot be neglected. We thus need an accurate model of the disk 
atmosphere for computing the spectral hardening 
factor~\citep{shimura,merloni,davis}. 
\item Additional effects to be considered in an accurate study are
the ones of limb darkening and of returning radiation.
\end{enumerate}


\begin{acknowledgments}
We would like to thank M. C. Miller and R. Takahashi
for critically reading a preliminary version
of this manuscript and providing useful feedback. E.B. would like to thank J. Brink and I. Mandel for pointing out a few typos
in the original Manko-Novikov metric. 
C.B. wishes to acknowledge the Horace Hearne Institute for 
Theoretical Physics at Louisiana State University and the 
Michigan Center for Theoretical Physics at the University of 
Michigan, for support and hospitality when this work was finalized.
The work of C.B. was supported by World Premier International 
Research Center Initiative (WPI Initiative), MEXT, Japan, and by 
the JSPS Grant-in-Aid for Young Scientists (B) No. 22740147.
E.B. acknowledges support from NSF Grants PHY-0903631, and would like to acknowledge
support and hospitality from the Institute for the Physics and Mathematics of the Universe at The University of Tokyo,
where this work was started.
\end{acknowledgments}


\appendix

\section{Manko-Novikov spacetimes}
\label{a-mn}
The Manko-Novikov metric is a stationary, axisymmetric, and 
asymptotically flat exact solution of the vacuum Einstein 
equations~\citep{m-n}. It is not a BH solution\footnote{The 
Manko-Novikov spacetimes have naked singularities and closed 
time-like curves exterior to a horizon. Therefore, the no-hair
theorem does not apply and the solution can have an infinite
number of free parameters. Let us notice, however, that all these
pathological features happen at very small radii and can be
neglected in our study, because they are inside the inner radius
of the disk. Here the basic idea is that naked singularities and
closed time-like curves do not exist in reality because they are ``covered'' by some exotic object,
whose \textit{exterior} gravitational field is described by the Manko-Novikov metric.}, but it 
can be used to describe the gravitational field outside a 
generic body like a compact star. The line element in 
quasi-cylindrical and prolate spheroidal coordinates is 
respectively
\be\label{eq-ds2}
ds^2 &=& - f \left(dt - \omega d\phi\right)^2
+ \frac{e^{2\gamma}}{f} \left(d\rho^2 + dz^2\right)
+ \frac{\rho^2}{f} d\phi^2 = \nonumber\\
&=& - f \left(dt - \omega d\phi\right)^2
+ \frac{k^2 e^{2\gamma}}{f}\left(x^2 - y^2\right)
\left(\frac{dx^2}{x^2 - 1} + \frac{dy^2}{1 - y^2}\right)
+ \frac{k^2}{f} \left(x^2 - 1\right)\left(1 - y^2\right) d\phi^2 \, ,
\nonumber\\
\ee
where
\be
f &=& e^{2\psi} A/B\, , \\
\omega &=& 2 k e^{- 2\psi} C A^{-1} 
- 4 k \alpha \left(1 - \alpha^2\right)^{-1} \, , \\
e^{2\gamma} &=& e^{2\gamma'}A \left(x^2 - 1\right)^{-1}
\left(1 - \alpha^2\right)^{-2} \, ,
\ee
and
\be
\psi &=& \sum_{n = 1}^{+\infty} \frac{\alpha_n P_n}{R^{n+1}} \, , \\\label{gammapdef}
\gamma' &=& \frac{1}{2} \ln\frac{x^2 - 1}{x^2 - y^2} 
+ \sum_{m,n = 1}^{+\infty} \frac{(m+1)(n+1) 
\alpha_m \alpha_n}{(m+n+2) R^{m+n+2}}
\left(P_{m+1} P_{n+1} - P_m P_n\right) + \nonumber\\
&& + \left[ \sum_{n=1}^{+\infty} \alpha_n 
\left((-1)^{n+1} - 1 + \sum_{k = 0}^{n}
\frac{x-y+(-1)^{n-k}(x+y)}{R^{k+1}}P_k \right) \right] \, , \\
A &=& (x^2 - 1)(1 + ab)^2 - (1 - y^2)(b - a)^2 \, , \\
B &=& [x + 1 + (x - 1)ab]^2 + [(1 + y)a + (1 - y)b]^2 \, , \\
C &=& (x^2 - 1)(1 + ab)[b - a - y(a + b)] 
+ (1 - y^2)(b - a)[1 + ab + x(1 - ab)] \, , \\\label{adef}
a &=& -\alpha \exp \left[\sum_{n=1}^{+\infty} 2\alpha_n 
\left(1 - \sum_{k = 0}^{n} \frac{(x - y)}{R^{k+1}} 
P_k\right)\right] \, , \\\label{bdef}
b &=& \alpha \exp \left[\sum_{n=1}^{+\infty} 2\alpha_n 
\left((-1)^n + \sum_{k = 0}^{n} \frac{(-1)^{n-k+1}(x + y)}{R^{k+1}} 
P_k\right)\right] \, .
\ee
Here $R = \sqrt{x^2 + y^2 - 1}$ and $P_n$ are the Legendre 
polynomials with argument $xy/R$:
\be
P_n &=& P_n\left(\frac{xy}{R}\right) \, , \qquad
P_n(x) = \frac{1}{2^n n!} \frac{d^n}{dx^n} \left(x^2 - 1\right)^n \, .
\ee
We notice that Eqs.~(\ref{gammapdef}), (\ref{adef}) and (\ref{bdef}) correct a few typos in the original Manko-Novikov metric
written in~\citet{m-n}: see~\citet{brink} and~\citet{thesis}.

The solution has an infinite number of free parameters:
$k$, $\alpha$, and $\alpha_n$ ($n=1, . . . , +\infty$). For 
$\alpha \neq 0$ and $\alpha_n = 0$, it reduces to the Kerr 
metric. For $\alpha = \alpha_n = 0$, we find the Schwarzschild 
solution. For $\alpha = 0$ and $\alpha_n \neq 0$, we obtain 
the static Weyl metric. Without loss of generality, we can
put $\alpha_1 = 0$ to bring the massive object to the origin
of the coordinate system. In this paper, we have restricted 
our attention to the subclass of Manko-Novikov spacetimes 
discussed in~\citet{gair}, where $\alpha_n = 0$ for $n \neq 2$.
Therefore, we have three free parameters ($k$, $\alpha$, and $\alpha_2$),
which are related to the mass, $M$, the dimensionless spin 
parameter, $a = J/M^2$, and the dimensionless anomalous 
quadrupole moment, $q = -(Q - Q_{\rm Kerr})/M^3$, of the object 
by the relations
\be
\alpha = \frac{\sqrt{1 - a^2} - 1}{a} \, , \qquad
k = M \frac{1 - \alpha^2}{1 + \alpha^2} \, , \qquad
\alpha_2 = q \frac{M^3}{k^3} \, .
\ee
Let us notice that $q$ measures the deviation from the quadrupole
moment of a Kerr BH. In particular, since 
$Q_{\rm Kerr} = - a^2 M^3$, the solution is oblate for 
$q > - a^2$ and prolate for $q < - a^2$. However, when
$q \neq 0$, even all the higher order multipole moments of the 
spacetime have a different value from the Kerr ones.

It is often useful to change coordinate system. The relation 
between the prolate spheroidal coordinates and the quasi-cylindrical
coordinates is given by
\be
\rho = k \sqrt{\left(x^2 - 1\right)\left(1 - y^2\right)} \, ,
\qquad
z = kxy \, ,
\ee
with inverse
\be
x &=& \frac{1}{2k} \left(\sqrt{\rho^2 + \left(z + k\right)^2} 
+ \sqrt{\rho^2 + \left(z - k\right)^2}\right) \, , \nonumber\\
y &=& \frac{1}{2k} \left(\sqrt{\rho^2 + \left(z + k\right)^2} 
- \sqrt{\rho^2 + \left(z - k\right)^2}\right) \, .
\ee
The relation between the standard Schwarzschild coordinates and
the quasi-cylindrical coordinates is given by
\be
\rho = \sqrt{r^2 - 2 M r + a^2 M^2} \sin\theta \, , \qquad
z = (r - M) \cos\theta \, .
\ee

\section{Circular orbits on the equatorial plane and ISCO}
\label{a-isco}
The line element of a generic stationary and axisymmetric
spacetime can be written as
\be
ds^2 = g_{tt} dt^2 + 2g_{t\phi}dt d\phi + g_{rr}dr^2 
+ g_{zz} dz^2 + g_{\phi\phi}d\phi^2 \, .
\ee
Since the metric is independent of the $t$ and $\phi$
coordinates, we have the conserved specific energy at
infinity, $E$, and the conserved $z$-component of the
specific angular momentum at infinity, $L_z$. This fact
allows to write the $t$- and $\phi$-component of the 
4-velocity of a test-particle as 
\be
u^t = \frac{E g_{\phi\phi} + L_z g_{t\phi}}{
g_{t\phi}^2 - g_{tt} g_{\phi\phi}} \, , \qquad 
u^\phi = - \frac{E g_{t\phi} + L_z g_{tt}}{
g_{t\phi}^2 - g_{tt} g_{\phi\phi}} \, .
\ee
From the conservation of the rest-mass, $g_{\mu\nu}u^\mu u^\nu = -1$,
we can write
\be
g_{rr}\dot{r}^2 + g_{zz}\dot{z}^2 = V_{\rm eff}(r,z) \, ,
\ee
where the effective potential $V_{\rm eff}$ for fixed $E$ and $L_z$
is given by
\be
V_{\rm eff} = \frac{E^2 g_{\phi\phi} 
+ 2 E L_z g_{t\phi} + L^2_z g_{tt}}{
g_{t\phi}^2 - g_{tt} g_{\phi\phi}} - 1 \, .
\ee
Writing the metric in quasi-cylindrical coordinates as 
in~(\ref{eq-ds2}), one finds
\be
V_{\rm eff} = \frac{E^2}{f} - \frac{f}{\rho^2}
\left(L_z - \omega E\right)^2 - 1 \, .
\ee
Circular orbits in the equatorial plane are located at the zeros
and the turning points of the effective potential: 
$\dot{r} = \dot{z} = 0$, which implies $V_{\rm eff} = 0$,
and $\ddot{r} = \ddot{z} = 0$, requiring respectively 
$\partial_r V_{\rm eff} = 0$ and $\partial_z V_{\rm eff} = 0$.
From these conditions, one can obtain the angular velocity,
$E$, and $L_z$ of the test-particle:
\be
\Omega_\pm &=& \frac{d\phi}{dt} = 
\frac{- \partial_r g_{t\phi} 
\pm \sqrt{\left(\partial_r g_{t\phi}\right)^2 
- \left(\partial_r g_{tt}\right) \left(\partial_r 
g_{\phi\phi}\right)}}{\partial_r g_{\phi\phi}} \, , \\
E &=& - \frac{g_{tt} + g_{t\phi}\Omega}{
\sqrt{-g_{tt} - 2g_{t\phi}\Omega - g_{\phi\phi}\Omega^2}} \, , \\
L_z &=& \frac{g_{t\phi} + g_{\phi\phi}\Omega}{
\sqrt{-g_{tt} - 2g_{t\phi}\Omega - g_{\phi\phi}\Omega^2}} \, ,
\ee
where the sign $+$ is for corotating orbits and the sign $-$
for counterrotating ones.
The orbits are stable under small perturbations if 
$\partial_r^2 V_{\rm eff} \le 0$ and $\partial_z^2 V_{\rm eff} \le 0$.
In Kerr spacetime, the second condition is always satisfied,
so one can deduce the radius of the innermost stable circular 
orbit (ISCO) from $\partial_r^2 V_{\rm eff} = 0$ . As first
noticed in~\citet{gair}, in general that is not true in 
Manko-Novikov spacetimes. For $q > 0$, $\partial_z^2 V_{\rm eff}$
is always smaller than zero and one finds that the ISCO
moves to larger radii as $q$ increases. When $q < 0$, for any value
of the spin parameter there are two critical values, say $q_1$ 
and $q_2$ with $q_1 > q_2$, and:
\begin{enumerate}
\item For $q > q_1$, one proceeds as in the case $q \ge 0$
and finds the radius of the ISCO through the equation
$\partial_r^2 V_{\rm eff} = 0$. $r_{\rm _{\rm ISCO}}$ decreases as
$q$ decreases.
\item For $q_2 < q < q_1$, there are two disconnected regions
with stable orbits. The standard region $r > r_1$, where $r_1$ 
is still given by $\partial_r^2 V_{\rm eff} = 0$, and an internal 
region $r_3 < r < r_2$, where $r_2$ is once again given by 
$\partial_r^2 V_{\rm eff} = 0$ (that is, circular orbits with 
$r_2 < r < r_1$ are radially unstable), while $r_3$ is given by 
$\partial_z^2 V_{\rm eff} = 0$ (that is, circular orbits with 
$r < r_3$ are vertically unstable). As $q$ decreases, $r_1$ 
decreases, while $r_2$ and $r_3$ increases. When $q = q_2$, 
$r_1 = r_2$.
\item For $q < q_2$, the ISCO is at $r_3$, which is given by 
$\partial_z^2 V_{\rm eff} = 0$. 
\end{enumerate} 
For $q > q_1$, the inner radius of the disk is the
radius of the ISCO, i.e. $r_{\rm in} = r_{\rm _{\rm ISCO}}$. For $q_2 < q < q_1$,
the energy and the angular momentum of the orbits in the 
region $r_3 < r < r_2$ are higher than the ones at $r > r_1$:
the result is that the inner radius of the disk is at $r_1$,
since the accreting matter reaches the orbit at $r_1$, which is 
a minimum of $E$ and $L_z$, and then plunges to the massive 
object. For $q < q_2$, the inner radius of the disk is $r_3$.

\begin{figure}
\par
\begin{center}
\includegraphics[type=pdf,ext=.pdf,read=.pdf,width=7.5cm]{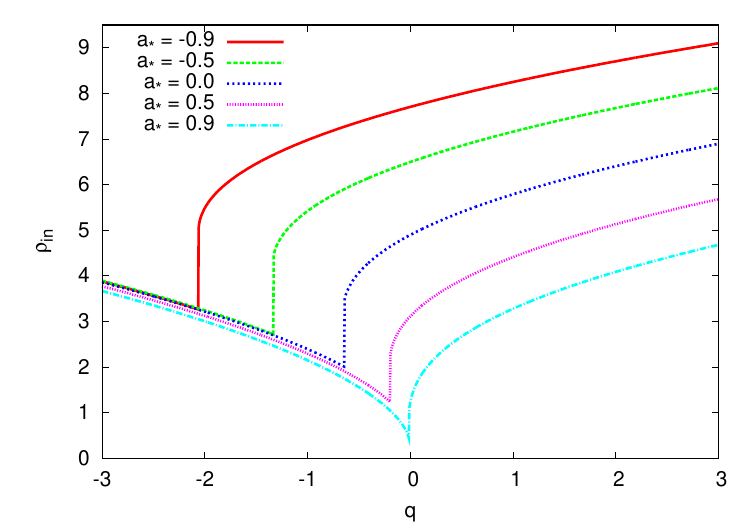}
\includegraphics[type=pdf,ext=.pdf,read=.pdf,width=7.5cm]{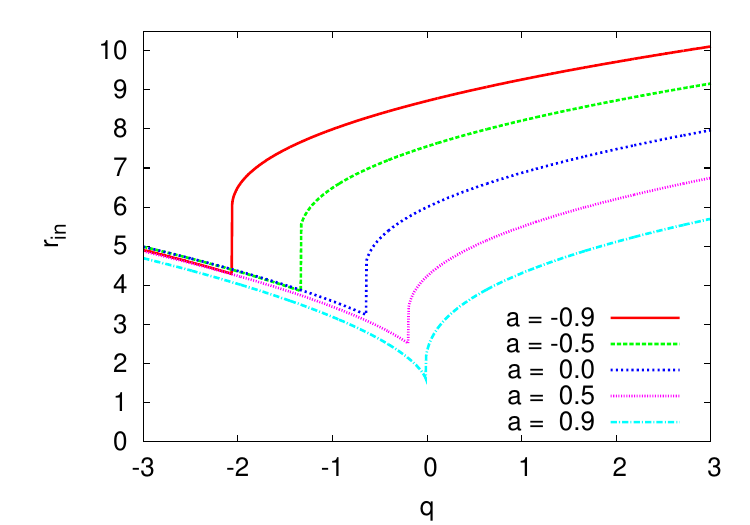}
\end{center}
\par
\caption{Radial coordinate of the inner radius of the disk
as a function of the anomalous quadrupole moment
$q$ for different values of the spin parameter $a$.
Left panel: radial coordinate in the quasi-cylindrical 
coordinates. Right panel: radial coordinate in the standard
Schwarzschild coordinates. Radial coordinate in unit $M = 1$.}
\label{f-isco}
\end{figure}

\begin{figure}
\par
\begin{center}
\includegraphics[type=pdf,ext=.pdf,read=.pdf,width=7.5cm]{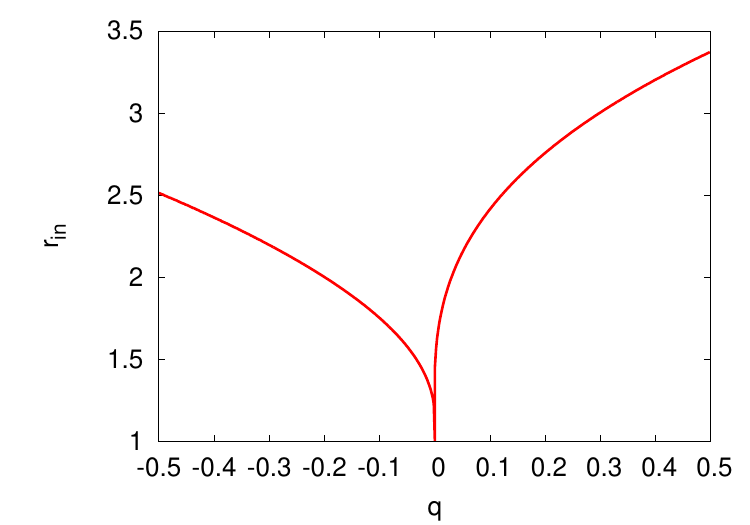}
\end{center}
\par
\caption{Radial coordinate in the standard Schwarzschild coordinates
of the inner radius of the disk as a function of 
the anomalous quadrupole moment $q$ for an extreme compact object 
$a \rar 1$. Radial coordinate in unit $M = 1$.}
\label{f-isco-extreme}
\end{figure}

In Fig.~\ref{f-isco}, we show the inner radius of the disk for a 
few values of the spin parameters $a$. In Fig.~\ref{f-isco-extreme},
we show the case of a maximally rotating object with $a \rar 1$:
it is remarkable that only in the case $q=0$ the inner radius of
the disk goes to $M$. For $q\neq0$, the inner radius of the disk is
significantly larger. Since current observations suggest that, at
least for some BH candidates, the inner radius of the disk
is consistent with the one of a fast-rotating Kerr BH,
deviations from the Kerr metric, if any, should not be large.

\end{document}